\documentstyle[aps,12pt,epsf]{revtex}
\begin{document}
\title{Exploring the nuclear pion dispersion relation through the anomalous coupling $\gamma\rightarrow \gamma^{\prime}\pi_0$}
\author{Alexander Volya, Kevin Haglin and Scott Pratt}
\address{National Superconducting Cyclotron Laboratory,
Michigan State University,\\
East Lansing, Michigan 48824-1321, USA}
\author{Vladimir Dmitriev}
\address{Budker Institute of Nuclear Physics,\\
Novosibirsk, 6300900 RUSSIA}
\date{\today}
\maketitle

\begin{abstract}
\baselineskip 14pt

We investigate the possibility of measuring the pion dispersion relation in
nuclear matter through the anomalous coupling in the reaction $\gamma
\rightarrow \gamma^{\prime}\pi_0$. It is shown that this reaction permits the
study of pionic modes for space-like momenta, $|{\bf k}_{\pi}|>\omega_\pi$. If
the pion is softened in nuclear matter due to mixing with the delta-hole state,
significant strength for this reaction is expected to move into the space-like
region. Competing background processes are evaluated, and it is concluded that
useful insight can be obtained experimentally, but only through a difficult
exclusive measurement.
\end{abstract}

\baselineskip 14pt
\section{Introduction}

Pions are expected to interact strongly with nuclear matter due to the mixture
of pionic and delta-hole states. The mixture is especially strong when the
momentum of the pionic mode approaches 300 MeV/c. In this momentum range the
energy of the pionic mode, $(k^{2}+m_{\pi }^{2})^{1/2}$, crosses the energy of
the delta-hole mode, $(k^{2}+M_{\Delta }^{2})^{1/2}-M_{N}$, and the energy of
the pionic branch should lower due to level repulsion. This topic drew a great
deal of attention in the late 1970s when it was thought that the pionic mode
might be pushed below zero energy at sufficient nuclear density, which would
result in pion condensation. An extensive review of pionic excitations and
condensation in nuclear matter was published by Migdal \cite{m78}, and
recently, kaon condensation has been the subject of several investigations
\cite{kn86}. Since such novel behavior requires higher densities, the
discussions are often in the context of neutron stars \cite{b93}, although
relativistic heavy ion collisions were also once proposed as a mechanism for
producing sufficient density for condensation.

Despite theoretical efforts in this area, experimental evidence of large
in-medium corrections to the pion dispersion relation is sparse. The most
promising information is from recent charge-exchange measurements from light
ions scattered off heavy nuclei. The cross-sections appear enhanced for
channels where pion-exchange is expected to dominate. This enhancement is
consistent with the lowering of the energy of the pionic branch, which reduces
the amount by which the exchanged pion is off-shell\cite{d94}. The
interpretation of this experiment suffers only from the fact that the probe is
hadronic and must traverse the surface of the heavy nucleus before
interacting. Heavy-ion collisions, which can produce matter at three to four
times nuclear density, were expected to create environments where the
dispersion relation was extremely distorted. However, experimental signals,
such as measuring pion spectra\cite{fs} or dilepton pairs\cite{gk87,kp90}, of
pionic properties in the interior of these regions proved difficult to extract.

In this paper we propose using high-energy photons to excite pionic
excitations. The anomalous coupling of a neutral pion to two photons, which is
responsible for the decay of the $\pi^{0}$, can be used to excite a pionic mode
in a heavy nucleus. The photon provides a clean probe for entering and exiting
the interior of the nucleus.  Unfortunately, this has the same drawback as the
charge-exchange experiment mentioned in the previous paragraph --- only
space-like excitations can be investigated. However, in-medium effects are
expected to lower the energy of the pionic branch, perhaps to the point that it
crosses into the space-like region, $|{\bf k}_{\pi}|>\omega_\pi$. Thus, the
reaction $\gamma \rightarrow \gamma^\prime + \pi^0 $, which is kinematically
forbidden in free space, is allowed in the nuclear medium. Furthermore, the
measurements of such a branch would provide direct evidence of in-medium
correction to the pion dispersion relation.

Our paper is organized into five parts. The next section briefly reviews the
in-medium corrections to the pion dispersion relation. The following section
shows the contribution to the cross section from the anomalous coupling to the
$\pi^{0}$, outlines the procedure one must follow to map out the pion
dispersion relation, and presents a discussion of how gauge-invariance
constrains the cross-section to disappear at the space-like-to-time-like
boundary. Due to this constraint, the contribution from the anomalous coupling
is reduced in the region of interest, which allows background processes to pose
a major problem. The bulk of the paper is comprised of estimates for background
processes which are presented in section \ref{section4}, where non-pionic
delta-hole channels are shown to provide most of the background. Given the
possibility for using free-electron lasers to create high-energy photon sources
with 100\% polarization, we also discuss the use of polarization measurements
to eliminate background. We find that the signal will only stand significantly
above the background if the final $\pi_0$ is detected in the final state with
an energy such that the target nucleus is in it's original ground state. We
conclude that an experimental investigation is challenging but tenable.

\section{The pion dispersion relation}

The delta-hole and nucleon-hole both contribute to the in-medium correction to
the self energy of the pion in nuclear matter. The coupling to the nucleon-hole
raises the energy of the pionic mode, while the coupling to the delta-hole
state significantly lowers the energy, especially for momenta approaching 300
MeV/c. Numerous theoretical works in this direction \cite{fs} have focused on
the correction to the pionic mode in perturbative
pictures\cite{weise_textbook,m78}. Sophisticated treatments of the pion self
energy were performed by Xia, Siemens and Soyeur\cite{xs94} and Korpa and
Malfliet\cite{korpa95}. There, self-consistent corrections to the delta's self
energy and width were included. We will forgo such lengthy calculations as we
wish to discuss how to experimentally observe the resulting dispersion
relation, rather than how to better calculate in-medium corrections.

Self energy corrections can be written diagrammatically as shown in Figure
\ref{diag1}. Vertex renormalization due to the effective four-point interaction
between deltas and nucleons is incorporated through the phenomenological
constant $g^{\prime}$. Nucleon-hole contributions can also be considered but
are much less important for ${\bf k}_{\pi}$ approaching 300 MeV/c.

Assuming an interaction Lagrangian density of the form,
\begin{equation}
{\cal L}_{\pi N\Delta}=
\frac{g_{\pi N\Delta}}{m_{\pi}}(\overline{\Delta}^{\mu}{\psi}
\partial_{\mu }\pi +h.c.)~,
\label{pindeltavertexeq}
\end{equation}
we obtain the self energy correction for the pion propagator shown in Figure
\ref{diag1}.

The delta propagator is assumed to have a Rarita-Schwinger form\cite{r73} for
all the calculations in this section, in which the width is inserted by
substitution $M_{\Delta} \rightarrow M_{\Delta}-i\Gamma_{\Delta}/2 $
\cite{jf95}.
\begin{equation}
G_{\Delta}^{\mu \nu}(q)=
\frac{i}{\not{q}-M_{\Delta}}\left[
-g^{\mu\nu}+\frac{1}{3}\gamma^{\mu}\gamma^{\nu}
+\frac{2}{3M_{\Delta}^2}q^\mu q^\nu
-\frac{1}{3M_{\Delta}}\left(q^\mu \gamma^\nu-\gamma^\mu q^\nu\right)
\right]
\end{equation}

The nucleon propagator in the presence of nucleons filled to the Fermi momentum
$p_{f}$ has a correction given by:
\begin{equation}
G_{N}(p)=\frac{i}{\not{p}-M_{N}+i\varepsilon }+2\pi (\not{p}-M_{N})\delta
(p_{0}^{2}-E^{2}(p))\Theta (p_{f}{\bf -|p|).} 
\end{equation}
With the obtained self-energy term a dispersion relation could be inferred by
finding a pole of the corresponding ``dressed'' pion propagator. Using these
equations and assuming the coupling strengths, $g_{\pi\Delta N}=2.0$ and
$g^{\prime}=0.8$, we obtain the pion dispersion relation presented in Figure
\ref{pion_disp_rel}.

The value of the effective in-medium coupling constant is not precisely known
although vertex corrections have been extensively studied. Many parameters used
in more sophisticated analyses\cite{xs94,korpa95} are still rather
uncertain. For instance the delta's mass and width as well as the nucleon's mass
in matter are somewhat uncertain, as are the coupling constants $g_{\pi
N\Delta}$ and $g^{\prime}$.

\section{Pion production through the anomalous coupling}

We consider production of a neutral pion through an effective photon ``decay''
inside nuclear matter as shown in Figure \ref{gpg}.

This process can only occur if the resulting pion is space-like,
\begin{equation}
k_{\pi }^{2}=(k_{i}-k_{f})^{2}=-2k_{i}\cdot k_{f}=-2\omega_{i}\omega_{f}
(1-\cos \theta)\leq 0\,,
\end{equation}
where $k_{i\text{ }}$and $k_{f}$ are the incoming and outgoing momenta of the
photon, and the outgoing photon leaves at an angle $\theta $. An interaction
that has a gauge invariant and parity conserving form is ${\cal L}_{\gamma
\gamma \pi }=\alpha /( 2\pi f_{\pi })F^{\mu \nu }\widetilde{F}_{\mu
\nu }\pi^0$, where $\alpha $ is the electromagnetic fine coupling constant, and
the pion decay constant is $f_{\pi }=93{\rm \,{MeV} }$. The transition
amplitude $\tau _{i\rightarrow f}=\left\langle f,\pi|{\cal L}_{\gamma \gamma
\pi}|i \right\rangle $ can be expressed in terms of the momenta and
polarizations of the incoming and outgoing photons.

\begin{eqnarray}
\tau _{i\rightarrow f} &=&\frac{\alpha }{2\pi f_{\pi }}2\epsilon _{\alpha
\beta \gamma \delta }k_{i}^{\alpha }k_{f}^{\beta }\chi _{i}^{\gamma }\chi
_{f}^{\delta }  \label{trans_amp_eq} \\
&=&\frac{\alpha }{2\pi f_{\pi }}({\bf k}{_{\pi}}^{2}-{\omega_{\pi}
}^{2})\sin (\alpha_i -\alpha_f )  \nonumber
\end{eqnarray}
where $k_{i}$ and $k_{f}$ are the initial and final momenta of the photon and
$\chi _{i}$ and $\chi _{f}$ are the corresponding polarizations. In the last
line of Eq. (\ref{trans_amp_eq}) the transition amplitude is written in terms
of the momentum of the pion, and the incoming and outgoing polarization angles,
$\alpha_i$ and $\alpha_f$, as measured relative to the reaction plane.  The
principle difficulty in obtaining the goals of this study comes from the fact
that the amplitude vanishes at the space-like-to-time-like boundary where the
square of a pion four-momentum is zero, which is precisely the kinematic region
of interest. This is a direct consequence of gauge symmetries involved in
coupling two photons to a pseudoscalar. This result may be problematic since we
investigate the area that is close to the space-like-to-time-like boundary, and
therefore the cross sections obtained are quite small. The rate of this
``decay'' can be expressed in terms of the matrix element $\tau$ \cite{iz80}:
\begin{equation}
d\Gamma =(2\pi )^4\delta ^4(k_{f}-p_{i}-k_{i})
|\tau _{i\rightarrow f}|^2 \frac{1}{2\omega _i}
\frac{d^3k_f}{2\omega _f(2\pi )^3}
\frac{d^3k_{\pi}}{2\omega _{\pi}(2\pi )^3}
\label{pi_rate}
\end{equation}
We will express our answer in terms of the energy and momentum transfer,
$\;\omega _{\pi }\equiv\omega _{i}-\omega _{f}$ and ${\bf k_{\pi }}\equiv{\bf
k_i-k_f}$ respectively. Substituting a Lorentzian form in place of the energy
preserving $\delta $ -function, allows the incorporation of a finite width to
the pionic state.
\begin{equation}
\delta (\omega -E)\rightarrow {\frac{1}{\pi }}{\frac{2\omega ^{2}\Gamma }{
(\omega ^{2}-E^{2})^{2}+\omega ^{2}\Gamma ^{2}}}  \label{eqn:six}
\end{equation}
The decay rate into a pionic mode of energy $\omega_\pi$ and momentum ${\bf k}_\pi$ can be expressed as:
\begin{equation}
\frac{d\Gamma }{dw_{\pi}d|{\bf k}_{\pi}|}=
\frac{\alpha ^2}{(2\pi )^{2}f_{\pi }^2}
\frac{|{\bf k}_{\pi}|({\bf k}_{\pi }^2-\omega _{\pi }^2)^2}
{2\omega _i^2}
\sin^2(\alpha_i -\alpha_f )
\frac{\omega _{\pi} \Gamma }
{(\omega _\pi -E_\pi )^2+\omega _{\pi }^2\Gamma^2}  
\label{total_pi_rate}
\end{equation}

To estimate the pion production cross section one must multiply this result by
the nuclear volume. Examples of scattering cross sections for the $ \gamma
\rightarrow \gamma^{\prime} +\pi^0$ process calculated for a Pb nucleus for
different incoming photon energies are shown in Figure \ref{pion_production_1},
where the on-shell energy of the pion is assumed to be $225MeV$ and momentum
transfer $|{\bf k}_{\pi}|$ is fixed at 275 MeV/c. Figure
\ref{pion_production_2} shows the dependence of the peak location and height
with respect to the on-shell energy. If the on-shell energy is not more then 25
MeV less than $|{\bf k}_\pi|$, the peak is probably too small to be observed.

The fact that the cross section is inversely proportional to the incoming
photon energy suggests the use of lower energies for a greater signal. Lower
energies also allow a more confident prediction of the background as very
massive and not well understood nucleonic resonances do not
contribute. Incoming photons with energies between 400 and 600 MeV are
satisfactory for our purposes given the above considerations and the
experimental ease with which they can be created. Lower-energy photons would be
difficult to deal with since the final photon could be confused with those from
nuclear processes such as giant-dipole decays.

For a 5 mm (a radiation length) lead target, one would need on the order of
$10^{14}$ photons to investigate the peak in the region of interest.  This
estimate was ascertained by requiring 100,000 final-state photons to correspond
to pion momenta between 225 and 275 MeV and pionic energies within 75 MeV of
the momentum. This number of photons is within the realm of current
experimental constraints, although the elimination of background to be
discussed in the next section will push the viability of these
measurements. The role of the width is important, as for vanishing widths the
shape of the differential cross section in Eq. (\ref{pi_rate}) becomes a sharp
spike which would be more easily observed. A confident calculation of the width
is not trivial, since the mixture of the pion and the delta-hole might
represent a significant contribution.

\section{Calculating background processes}
\label{section4}
We consider two processes that might overwhelm the $\gamma \rightarrow
\gamma^{\prime} +\pi^0$ reaction, namely ordinary Compton scattering, that
could also proceed through an intermediate $\Delta $ resonance, see Figure
\ref{compton_feynmann}, and reactions that produce a $\Delta $-hole (which
decays into $\pi N$) as a final state, see Figure \ref{d_diagrams}. These
reactions are the same order in $\alpha $ as $\gamma \rightarrow
\gamma^{\prime} +\pi ^0$. One might expect these background processes to be
smaller than the simple $\pi_0$ production process since they are further
off-shell in the kinematic region we are exploring. However, even though
$\gamma \rightarrow \gamma^{\prime} +\pi^0$ might be nearly on-shell, a gauge
constraint forces the matrix element to zero at the $k_{\pi }^{2}=0$ boundary
and allows other channels to compete. At the end of this section we will
demonstrate that background processes overwhelm the signal, but that by gating
on a final-state $\pi_0$, one might sufficiently suppress the background. We
also report on our investigation of using polarization measurements to project
the signal from the background. Unfortunately, our estimate of the background
processes is done without the benefit of an experimental measurement of
relevant processes in vacuum, e.g. $\gamma + p \rightarrow p^{\prime} +
\gamma^{\prime} +\pi^0$. Certainly such measurements are possible and would
greatly increase the confidence with which we present the background.

First, we discuss our estimate of normal Compton scattering as illustrated in
Figure \ref{compton_feynmann}. We consider the intermediate state to be either
a delta or a proton. The diagrams with the delta as an intermediate state
provide the dominant contribution. The photon is assumed to interact with the
baryon via the following couplings \cite{tw78,js73}:
\begin{eqnarray}
{\cal L}_{p\gamma p}&=&e\overline{\psi}\gamma_\mu A^\mu \psi
\nonumber\\
{\cal L}_{\gamma N\Delta}&=&e\left\{ 
\frac{G_1}{M_N}\overline{\Delta}^\mu \gamma^\nu \gamma_5 \psi F_{\mu \nu} 
+\frac{G_2}{M_N^2}\partial^\nu \overline{\Delta}^\mu \gamma_5\psi F_{\mu \nu}
+h.c.
\right\}
\label{photoexeq}
\end{eqnarray}
We observed that the contribution from the second term proportional to $G_2$ is
small, and we neglected it in our analyses.The value of $G_1=2.63$ has been
determined from experiments\cite{jn94,gg93}. The regular Compton process was
also small, and thus did not warrant including more sophisticated coupling,
e.g. through magnetic moments.

The result for the cross section is shown in Figure
\ref{compton_results}. Again, we have assumed that the photon lost a momentum
$|{\bf k_{\pi}}|=275$ MeV/c. The form would be a sharp peak at low energy if it
were not for our replacing the final delta function in the cross-section by a
Lorentzian, giving the nucleon an effective width of 25 MeV. Since an on-shell
nucleon with 275 MeV/c of momentum has only 50 MeV of energy, there is little
contribution in the kinematic region of interest, where $\omega_\pi$ approaches
$|{\bf k_{\pi} }|$. Since the kinematics were effectively smeared by the Fermi
motion, changing the nucleon's width had little effect.

The primary contribution to the background derives from production of a
delta-hole in the final state. Since this is precisely the process that mixes
with the pion due to its kinematic proximity to the pionic mode, it is not
surprising. Even though the delta-hole should be 50 MeV in energy higher than
the pionic branch of the dispersion relation, its contribution is not
constrained to go to zero when $|{\bf k_{\pi}}|$ equals $\omega_\pi$. This lack
of a constraint derives from the fact that a spin 3/2 delta and a spin 1/2
nucleon-hole do not necessarily form a pseudo-vector, and couple to $\partial
\pi$, but can also couple to $J=2$. Furthermore, the delta will decay into a
nucleon and a pion. If the pion is charged, it can radiate photons readily
since it is light and moves quickly. We find that only by gating on the
presence of the $\pi_0$ and by requiring the target nucleus to be left in it's
ground state, can one confidently translate the measurement into information
regarding the nuclear pion disperion relation.

The diagrams used for calculating the contributions for a delta-hole being in
the final state are shown in Figure \ref{d_diagrams}. In order to maintain
gauge invariance, the decay of the delta into pions is included. This decay is
also crucial as Bremsstrahlung off the light charged pion is important. The
coupling of the electromagnetic field to the delta is accomplished via minimal
substitution. 

The Lagangian for the delta that results in the Rarita-Schwinger form of the
propagator is:
\begin{equation}
{\cal L}_\Delta = -\overline{\Delta}^\alpha \left\{
(i\not{\partial}-M_\Delta )g_{\alpha \beta}
-(\gamma_\alpha i\partial_\beta +i\partial_\alpha\gamma_\beta )
+\gamma_\alpha i\not{\partial} \gamma_\beta
+M_\Delta \gamma_\alpha\gamma_\beta
\right\}
\Delta^\beta
\end{equation}
The associated coupling of a delta to a photon, through minimal substitution, is:
\begin{equation}
{\cal L}_{\Delta\Delta\gamma}=e\overline{\Delta^\alpha}\left\{
g_{\alpha\beta}\gamma^\mu - g_\alpha^\mu \gamma_\beta 
-g_\beta^\mu \gamma_\alpha + \gamma_\alpha\gamma^\mu\gamma_\beta
\right\}\Delta^\beta A_\mu
\end{equation}

Due to the derivative nature of the $\pi N\Delta$ coupling shown in
Eq. (\ref{pindeltavertexeq}), minimal substitution requires a four point
coupling of the photon to the $\pi N\Delta$ vertex.
\begin{equation}
{\cal L}_{\gamma\pi N\Delta}=ie\frac{g_{\pi N\Delta}}{m_\pi}\Delta^\mu
A_\mu \psi \pi +h.c.
\end{equation}
Minimal substitution from the interaction term in Eq. (\ref{photoexeq}), also
results in a four-point coupling, which we will neglect since $G_2$ is set to
zero.

The 20 terms necessary for creating a nucleon, a photon and a pion in the final
state are shown in Figure \ref{d_diagrams}. A fixed width was used for the
delta, which although is not realistic, simplified self-consistency checks
regarding gauge invariance and the Ward-Takahashi identity. Since the
delta-hole is not far off-shell, the answer should not vary greatly by
incorporating an energy and density dependent width.

Transition elements were calculated for specific linear polarizations,
$\alpha_i$ and $\alpha_f$ of the incoming and outgoing photons. The scattering
plane of the photons defines the angles. For our signal, $\gamma \rightarrow
\pi_0 \gamma^{\prime}$, the dependence is proportional to
$\sin^2(\alpha_i-\alpha_f)$ as shown in Eq. (\ref{trans_amp_eq}). Since the
final states illustrated in Figure \ref{d_diagrams} are three-body states with
large widths from the delta decay, the final three particles could be assumed
to be on-shell. Numerical integrations were performed over all final-state
variables except for the photon polarizations and the energy and momentum lost
by the photon, $\omega_{\pi}$ and $|{\bf k}_\pi |$.

The cross sections for the background are shown in Figure \ref{bg_delta}. The
upper graph shows the background for charged pions, while the
lower graph presents the background for the case where a neutral pion is
created. The creation of a charged pion overwhelms the signal by more than an
order of magnitude. This strength comes from the radiation off the light,
fast-moving, charged particle.

Polarization does not significantly ameliorate the background as shown in
Figure \ref{bg_delta}. The signal is proportional to
$\sin^2(\alpha_i-\alpha_f)$ which has the same shape as part of the spin-2
delta-hole state.

By gating on neutral pions, one can reduce the background by an order of
magnitude as shown by comparing the upper and lower panels of Figure
\ref{bg_delta}. Although this background is now comparable to the signal, it is
still possible that a charged pion could be created and undergo a charge
exchange with the medium, resulting in a neutral pion. 

For the reasons discussed above, the best hope of extracting the signal would
be to experimentally verify that the nucleus remained in it's ground
state. This could be accomplished by observing the outgoing $\pi_0$ with an
energy such that the nucleus would be constrained to it's ground state. This
effectively eliminates both the delta-hole and nucleon-hole background from
consideration. If the incoming nucleon in Figure \ref{d_diagrams} returns to
it's original state within the Fermi sea, such a process can be considered as a
contribution to the pionic state, rather than as a separate background process.

This exclusive process of creating a $\pi_0$ without exciting the nucleus would
certainly reduce the overall size of the signal shown in Figures
\ref{pion_production_1} and \ref{pion_production_2}. The magnitude of this
reduction can be considered as the probability for the pion to escape the
nucleus. Careful estimates of a pion's width in the medium have been performed,
with the conclusions that the width is in the 10-30 MeV range for pions with
momenta less than 300 MeV/c\cite{korpa95}. Using the expression for the mean free path, $\lambda=v/\Gamma$, one can expect that the mean free path is on the
order of two to five Fermi.

An estimate of the escape probability is shown in Figure \ref{escape_prob} as a
function of the mean free path $\lambda$. The escape probability is the
average, $\left<\exp{(-x/\lambda)}\right>$, where the average is performed over
all originating points within a 7.0 Fermi sphere, and over all possible
directions, and $x$ is the distance the pion must travel to escape the
sphere. By viewing Figure \ref{escape_prob} one can see that absorption will
reduce the signal by approximately an order of magnitude. Combined with the
difficulty in detecting all three photons (two from the decay of the $\pi_0$),
the exclusive measurement becomes difficult. If lead is the target, the
first excited state is at 2.6 MeV, meaning that the photons must be measured to
better than 1.0 MeV to eliminate all excited states from consideration. Of
course, even the reduction of the number of excited states to a small number
would greatly reduce the unwanted delta-hole and nucleon-hole backgrounds. 

\section{Conclusions}
The subject of in-medium hadron masses has been historically inconclusive. The
subject of heavier mesons such as the rho, is perhaps of even greater interest
due to the connection to chiral symmetry restoration. The pion's in-medium
properties are unique in that the dispersion relation may dip down into the
space-like region at normal nuclear density. This permits the use of simple
scattering experiments, such as the process proposed here, to investigate
pionic modes.

The prospects for investigating the dispersion relation hinge on detection of 
neutral pions to eliminate the background contributions. Although detection of
the neutral pion can be accomplished with the same apparatus used to measure
the scattered photon, the reduction in statistics might make the experiment
untenable. It should be pointed out that theoretical efforts can be improved,
especially for the understanding of finite-size effects for the pionic modes in
the nucleus. An optical model calculation would be welcome.

Although the experimental challenges are significant, this line of
investigation does offer the possibility for definitive evidence in the search
for in-medium hadronic properties. The implications of such evidence would
extend beyond the context of finite nuclei, to infinite nuclear matter and
neutron stars.

\acknowledgments{This work was supported by the National Science Foundation,
grant no.  PHY-9513900.}

\begin{figure}
\begin{center}
\end{center}
\caption{Delta-hole contributions to the pion self energy.
\label{diag1}}
\end{figure}

\begin{figure}
\begin{center}
\end{center}
\caption{The pion dispersion relation is shown for both the vacuum case (solid
line) and with an effective coupling $g_{\pi N\Delta}$=2.0 and $g^{\prime}$=0.8
(dashed line) in Figure \ref{diag1}.
The thin strait line shows the space-like-to-time-like boundary.
\label{pion_disp_rel}}
\end{figure}

\begin{figure}
\begin{center}
\end{center}
\caption{Pion production diagram.
\label{gpg}}
\end{figure}

\begin{figure}
\begin{center}
\end{center}
\caption{Neutral pion production through the anomalous coupling $\gamma
\rightarrow \gamma^{\prime} +\pi^0$. The on-shell energy is 225 MeV, the
momentum transfer $|{\bf k}_{\pi}|$ is 275 MeV/c, and the pion's width is 50
MeV. Cross sections are shown for three incoming photon energies: 500 MeV
(solid line), 1.0 GeV (long dashes) and 2.0 GeV (short dashes).
\label{pion_production_1}}
\end{figure}

\begin{figure}
\begin{center}
\end{center}
\caption{Neutral pion production through the anomalous coupling $\gamma
\rightarrow \gamma^{\prime} +\pi^0$. On-shell energies are 225 MeV (solid
line), 250 MeV (long dashes) and 275 MeV (short dashes).  The momentum transfer
$|{\bf k}_{\pi}|$ is 275 MeV/c, and the pion width is 50 MeV.  The incoming
photon has an energy of 500 MeV.
\label{pion_production_2}}
\end{figure}

\begin{figure}
\begin{center}
\end{center}
\caption{Feynman diagrams for the background Compton-like processes $\gamma
+N\rightarrow \gamma^{\prime} +N^{\prime}$.
\label{compton_feynmann}}

\end{figure}

\begin{figure}
\begin{center}
\caption{The contribution from $\gamma +N \rightarrow
\gamma^{\prime}+N^{\prime}$ assuming the outgoing nucleon has a width of 25
MeV. This component is negligible for energy transfers greater than 200 MeV.
\label{compton_results}}
\end{center}
\end{figure}

\begin{figure}
\begin{center}
\caption{Matrix elements for calculating background processes where a
pion, photon and excited nucleon are in the final state. These diagrams can be
thought of as processes where a delta-hole (which subsequently decays) is
created.
\label{d_diagrams}}
\end{center}
\end{figure}

\begin{figure}
\begin{center}
\caption{The cross-section for the background process where a charged pion is
created is shown in the upper panel. By requiring the created pion to be
neutral, the background is greatly reduced as shown in the lower
panel. Polarization projections are shown for $\alpha_i=0,~\alpha_f=0$ (solid
line),$\alpha_i=0, ~\alpha_f=\pi /2$ (short dashes),
$\alpha_i=\pi/2,~\alpha_f=0$ (long dashes), $\alpha_i=\pi /2,~\alpha_f=0$
(dot-dashes).
\label{bg_delta}}
\end{center}
\end{figure}

\begin{figure}
\begin{center}
\caption{The escape probability of a pion from a nucleus with the radius of 7
Fermi as a function of the mean free path $\lambda$.
\label{escape_prob}}
\end{center}
\end{figure}

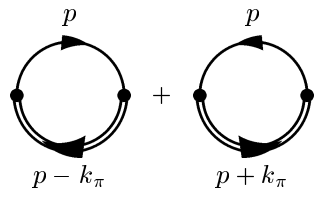
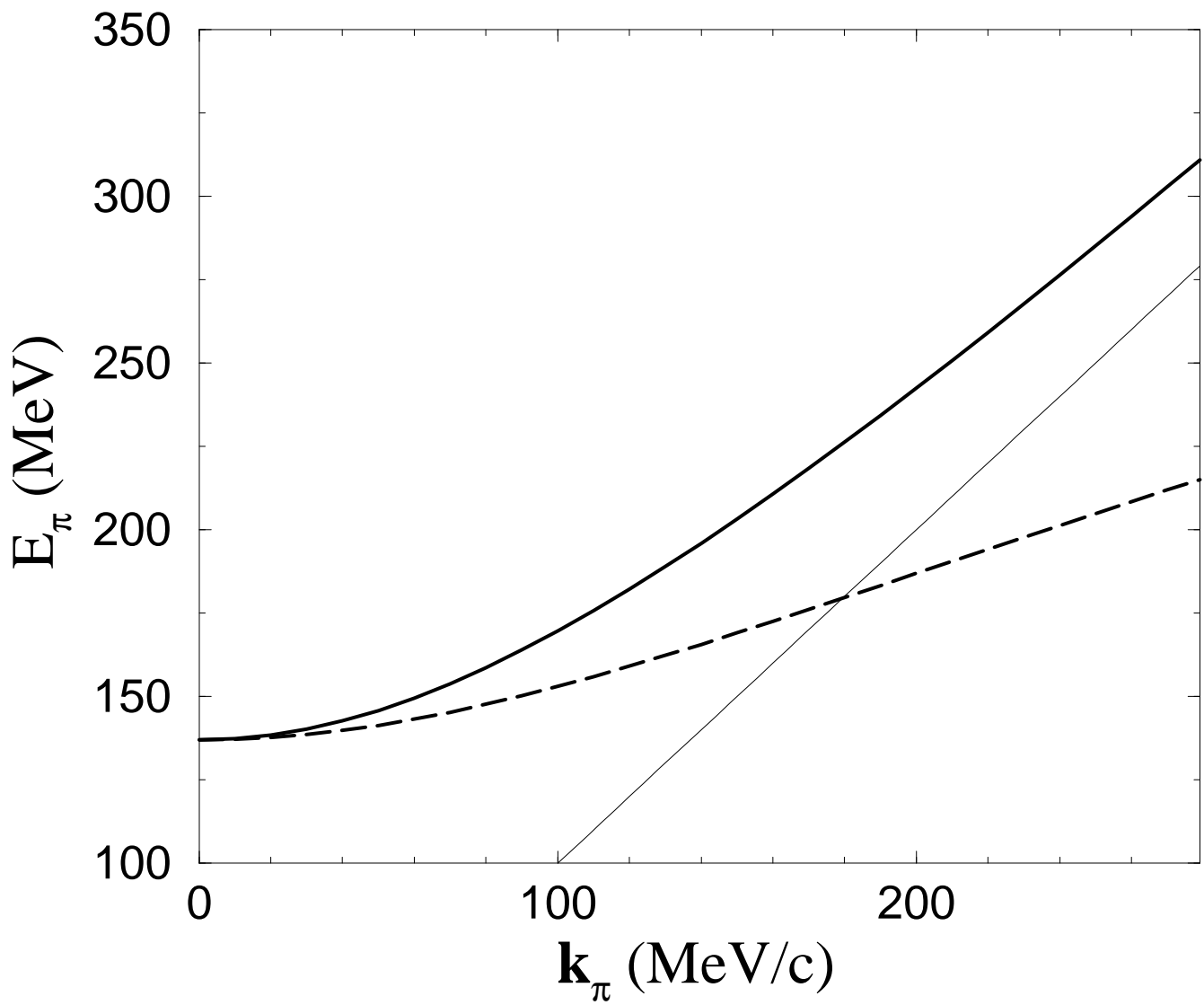
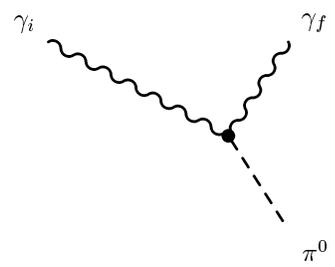
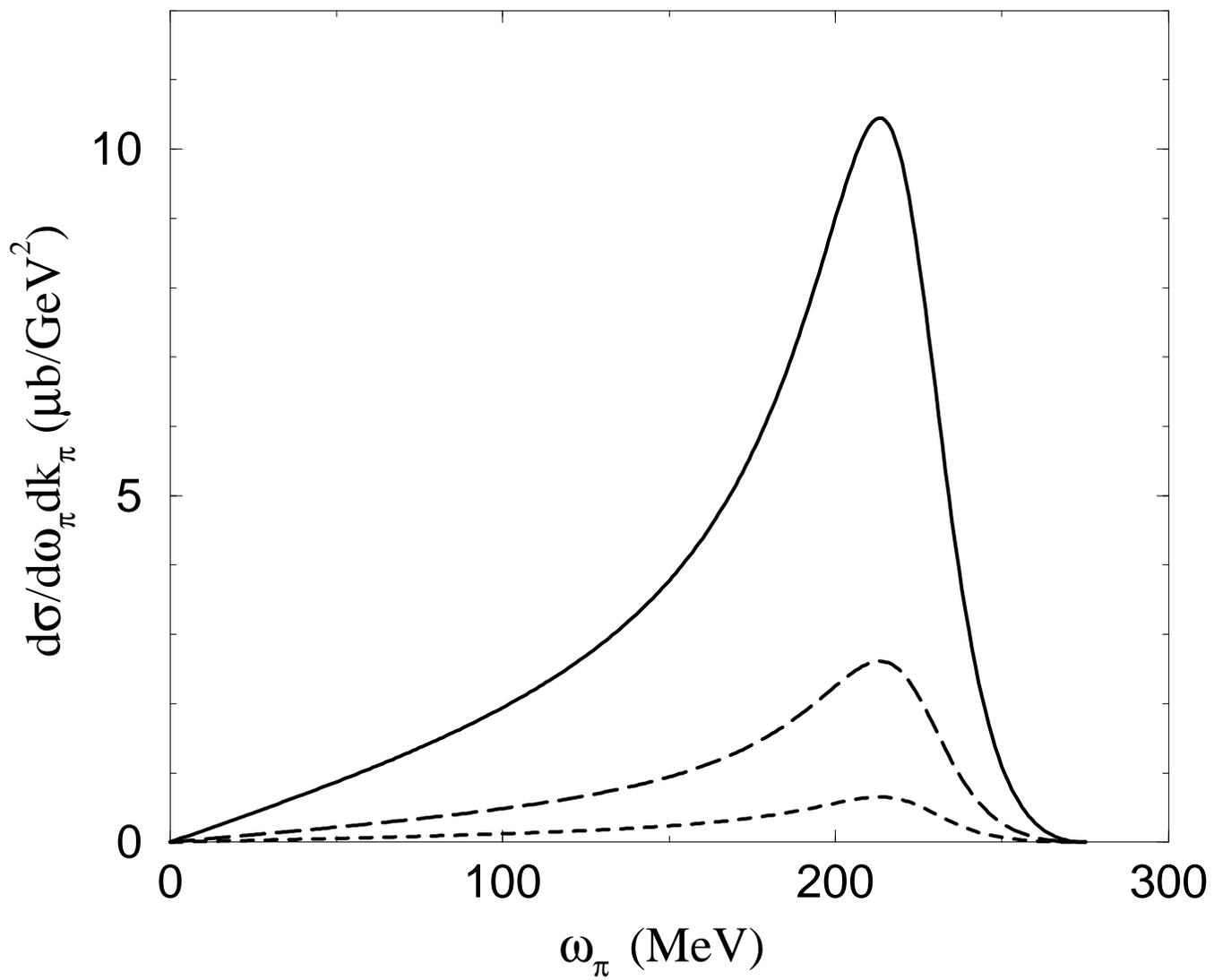
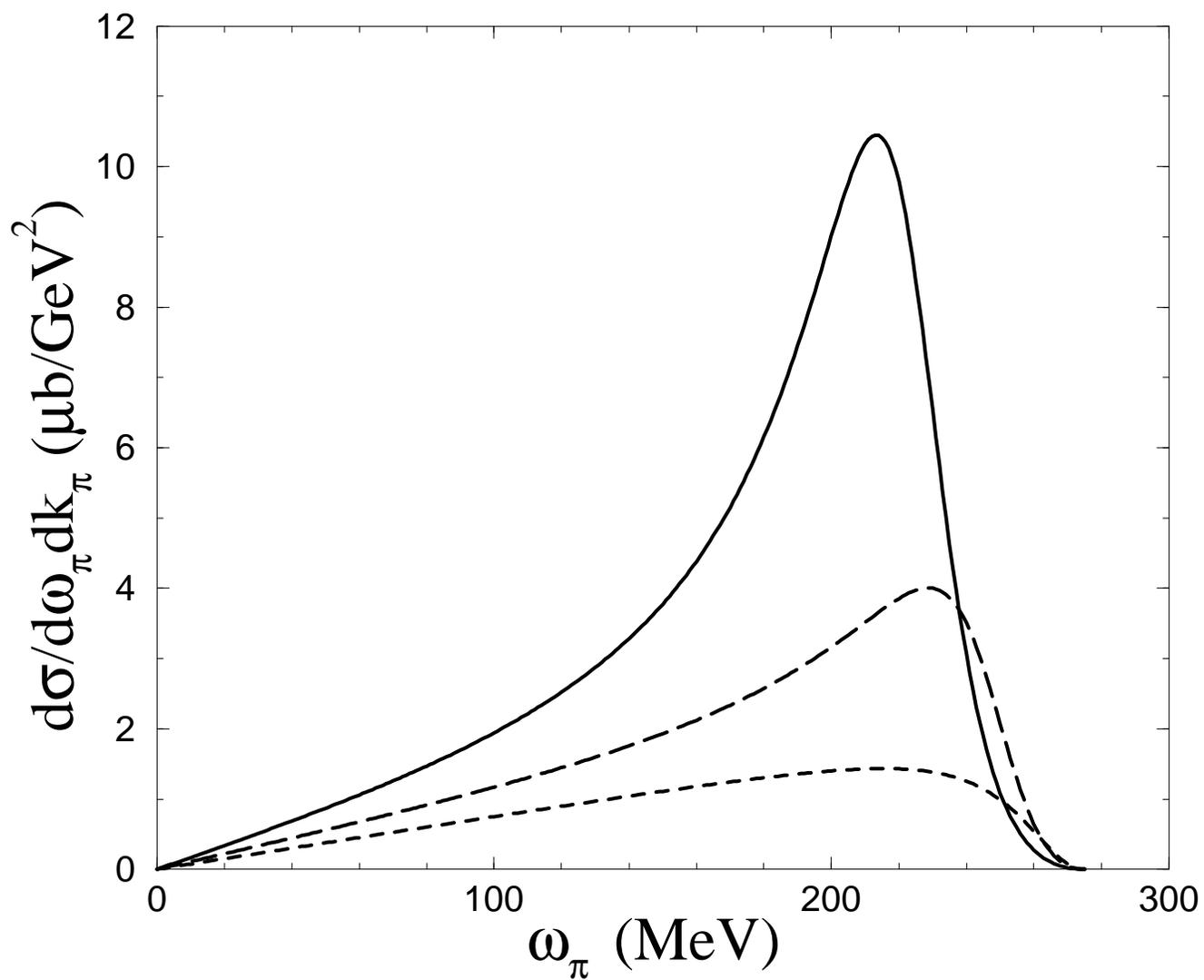
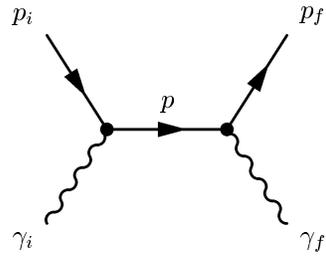
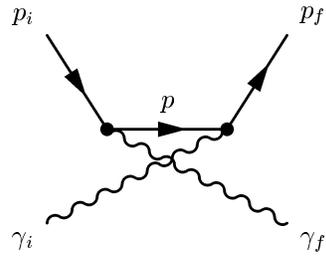
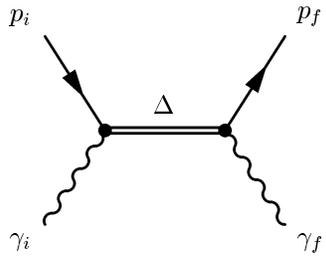
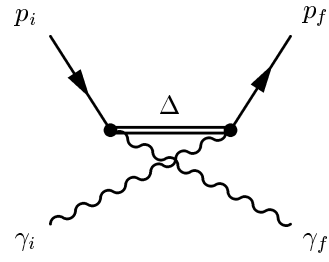
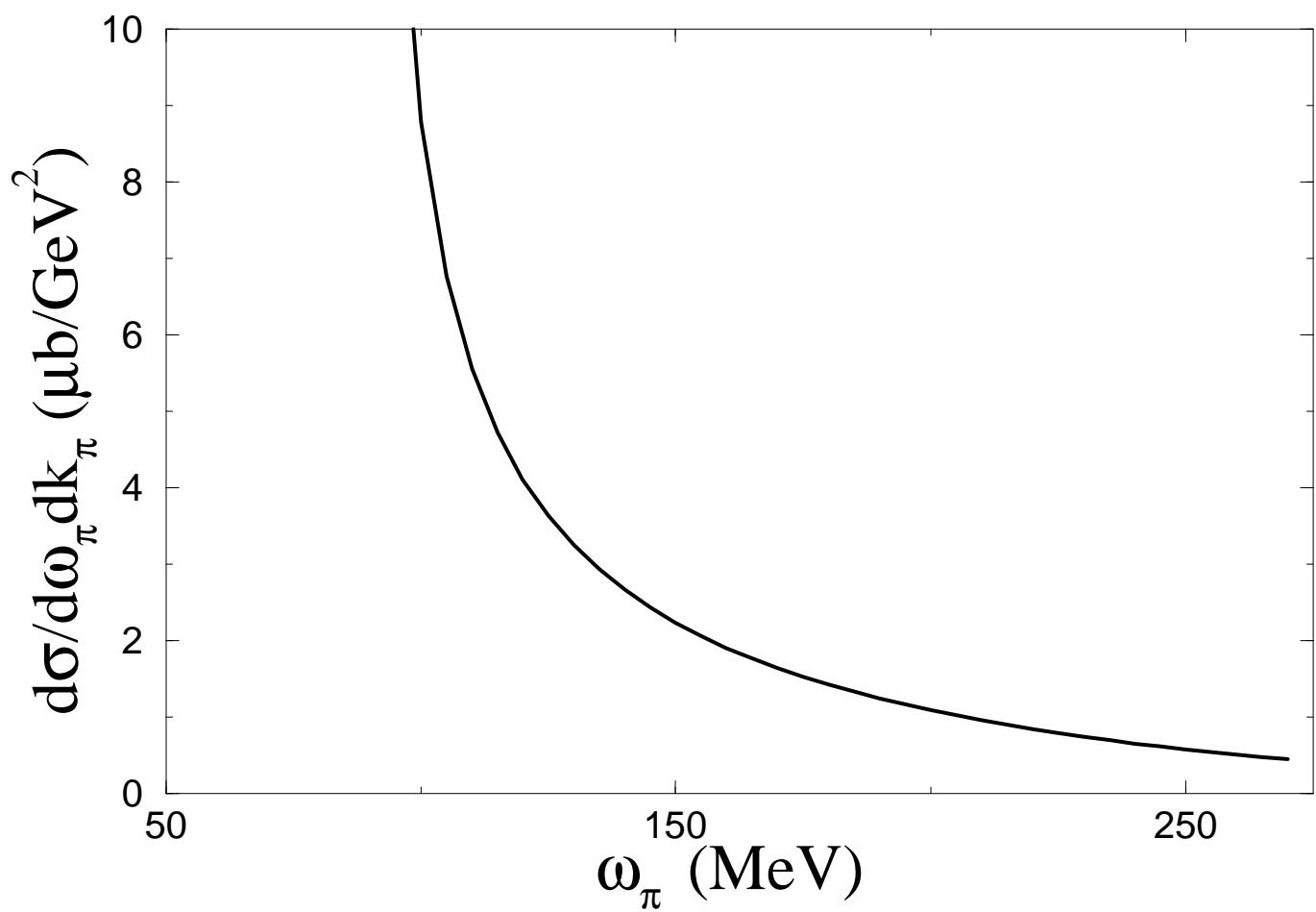
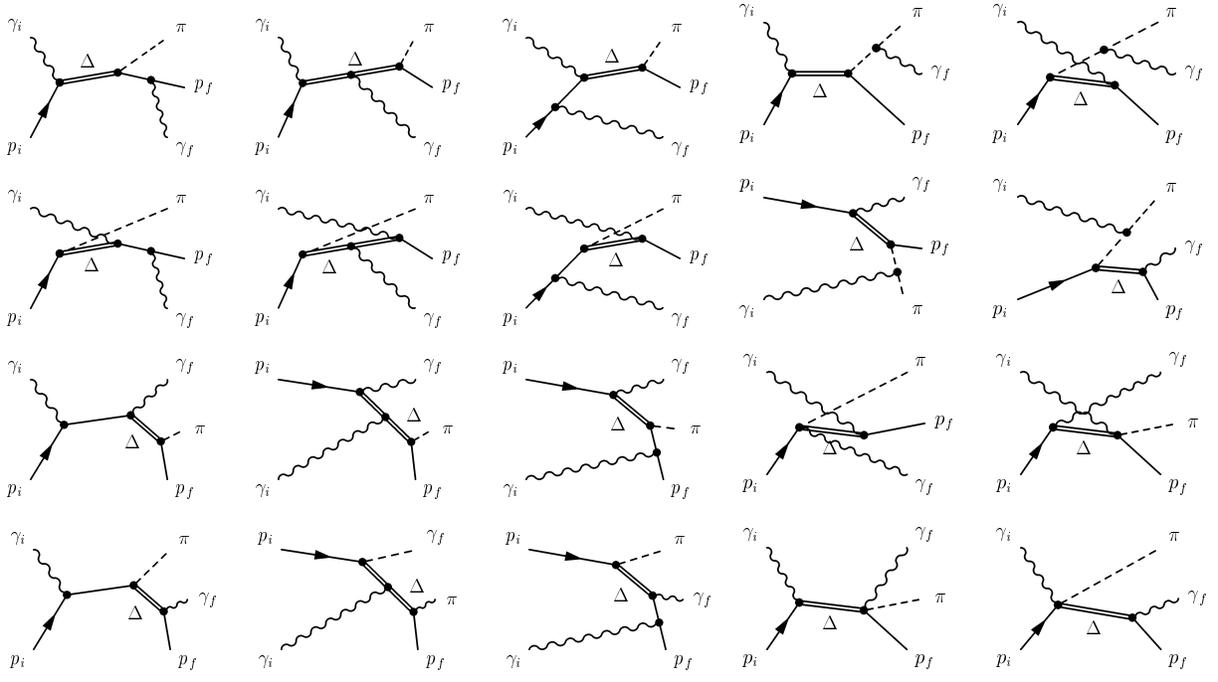
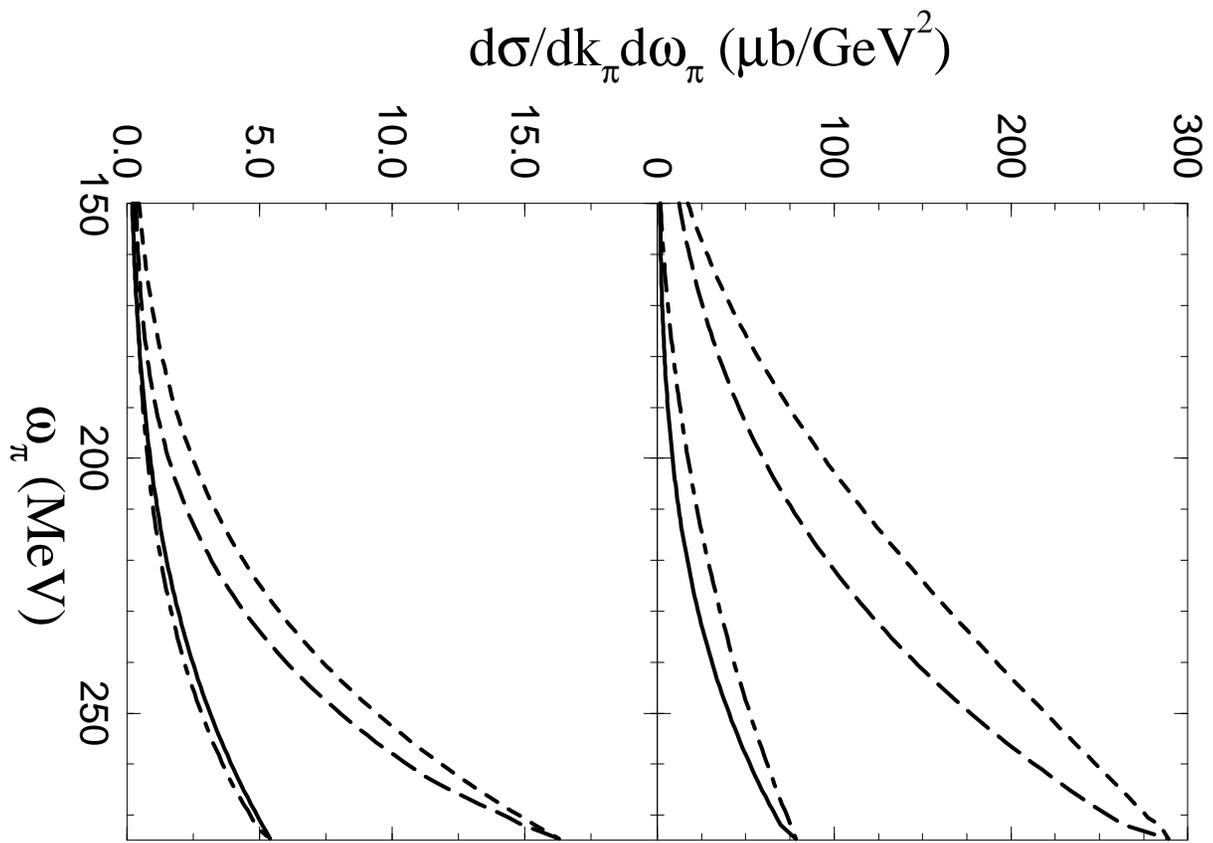
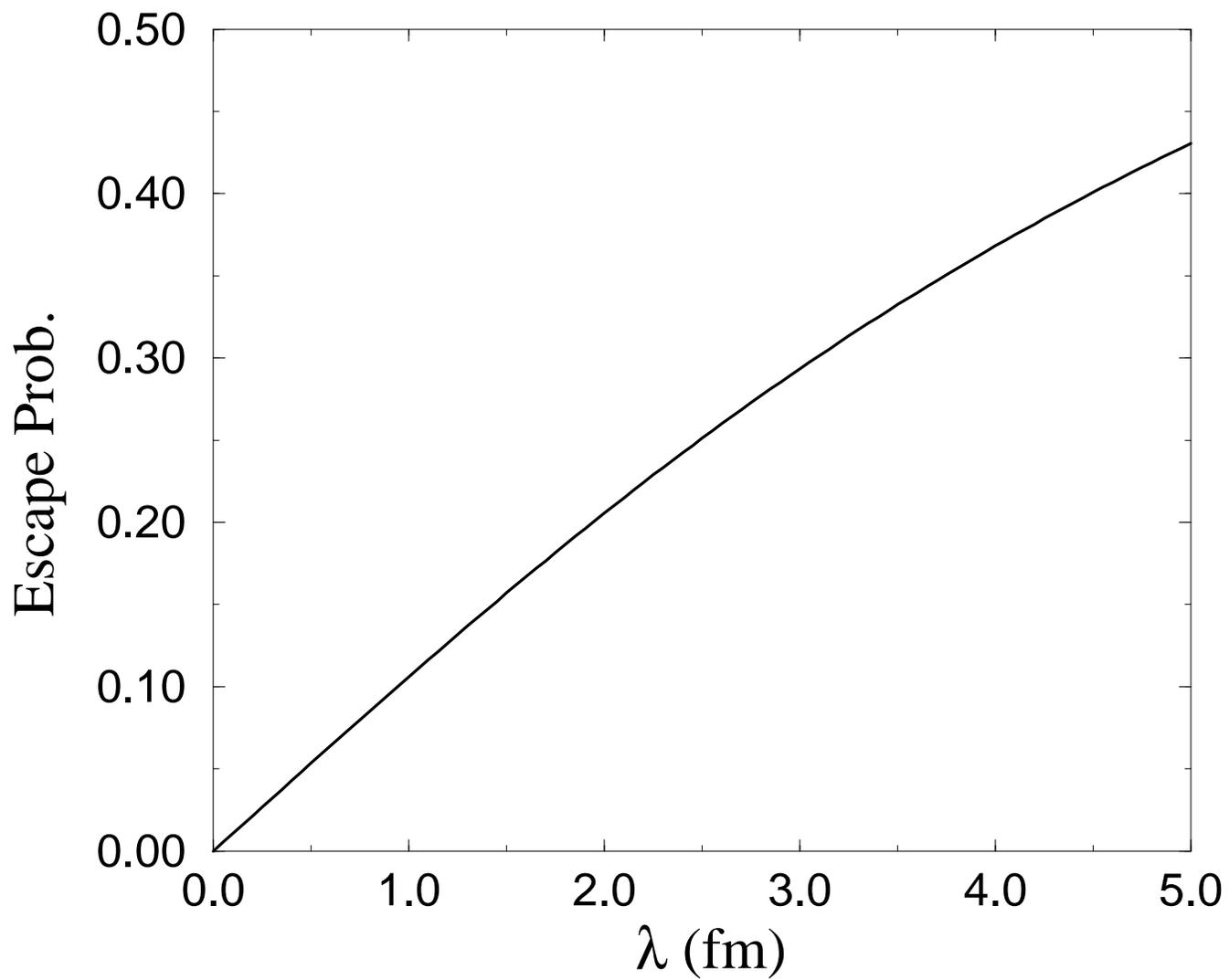
\end{document}